\documentclass{article}\usepackage{spconf,amsmath,graphicx,cite}     

\usepackage{amsfonts}
\usepackage{graphics} 
\usepackage[tight,footnotesize]{subfigure}

\def\twon #1{\left\|#1\right\|_2}
\def\onen #1{\left\|#1\right\|_1}

\def\sgn #1{\text{sgn}#1}
\def\abs #1{\left|#1\right|}
\def\inp #1{\left\langle#1\right\rangle}

\def\st{\textrm{ subject to }}

\def\bI{\mathbb{I}}

\def\bS{\mathbb{S}}

\def\m #1{\boldsymbol{#1}}

\def\cF{\mathcal{F}}

\def\cL{\mathcal{L}}

\def\cP{\mathcal{P}}

\def\bee{\begin{equation}}
\def\ene{\end{equation}}

\def\beq{\begin{eqnarray}}
\def\enq{\end{eqnarray}}
\def\lentwo{\setlength\arraycolsep{2pt}}

\newtheorem{thm}{Theorem}

\newcommand{\BOX}{\hfill\rule{2mm}{2mm}}

\def\equ #1{\begin{equation}#1\end{equation}}
\def\equa #1{\begin{eqnarray}#1\end{eqnarray}}
\def\sbra #1{\left(#1\right)}

\def\lbra #1{\left\{#1\right\}}

\def\st {\text{ subject to }}

\title{Robust Compressive Phase Retrieval via L1 Minimization With Application to Image Reconstruction}
\name{Zai Yang$^*$, Cishen Zhang$^\dag$, and Lihua Xie$^*$, Fellow, IEEE}
\address{$^*$EXQUISITUS, Centre for E-City, School of Electrical and Electronic Engineering,\\
Nanyang Technological University, 639798, Singapore\\
$^\dag$Faculty of Engineering and Industrial Sciences, Swinburne University of Technology, \\Hawthorn VIC 3122, Australia}
\begin{document}
\maketitle

\begin{abstract} Phase retrieval refers to a classical nonconvex problem of recovering a signal from its Fourier magnitude measurements. Inspired by the compressed sensing technique, signal sparsity is exploited in recent studies of phase retrieval to reduce the required number of measurements, known as compressive phase retrieval (CPR). In this paper, $\ell_1$ minimization problems are formulated for CPR to exploit the signal sparsity and alternating direction algorithms are presented for problem solving. For real-valued, nonnegative image reconstruction, the image of interest is shown to be an optimal solution of the formulated $\ell_1$ minimization in the noise free case. Numerical simulations demonstrate that the proposed approach is fast, accurate and robust to measurements noises.
\end{abstract}



\section{Introduction}
Many imaging techniques reconstruct a signal from its frequency or Fourier measurements. But in practice the phase information of the data in the frequency domain may not be available to the detecting and sensing devices, e.g., in X-ray crystallography \cite{millane1990phase}. It therefore arises the problem of recovering a signal from its Fourier magnitude only measurements, known as phase retrieval. Other applications of phase retrieval include optics \cite{millane1990phase}, diffraction imaging \cite{bunk2007diffractive}, astronomical imaging \cite{fienup1987phase} and magnetic resonance imaging \cite{yang2013sparse}, to name just a few. Since more than one signal can result in the same Fourier magnitude measurements, the phase retrieval problem is ill-posed. For example, a solution to the problem can be subject to so-called global phase ambiguity including a constant global phase, spatial shift and conjugate inversion. Though such ambiguities are acceptable in practical applications, there may exist infinitely many solutions beyond these trivial associates. To resolve the problem, it is shown in \cite{hayes1982reconstruction} that twofold oversampling in each dimension in the frequency domain almost always specifies a unique solution (up to global phase) for finitely supported, real-valued and nonnegative signals while it does not work for 1D signals. Due to the nonlinear relationship between the magnitude measurements and the signal of interest, the algorithm design remains a major problem in phase retrieval. Current approach to signal reconstruction with the oversampling technique include the popular iterative projection algorithms \cite{luke2005relaxed} and recent convex optimization method \cite{candes2011phaselift}.


It is of great interest to reduce the required number of measurements in phase retrieval since every single one of the measurements may have a potential time and/or power cost. Moreover, oversampling can be inconvenient or impossible in applications, e.g., Bragg sampling from
periodic crystalline structures \cite{marchesini2008ab}. Compressed sensing (CS) \cite{candes2006near} is an emerging technique which aims at reconstructing a high dimensional signal from its low dimensional linear measurements under a sparsity prior. A prosperous direction is to combine CS with phase retrieval, known as compressive phase retrieval (CPR) \cite{moravec2007compressive,marchesini2008ab}, which reduces the required number of measurements by exploiting the signal sparsity. It is shown in  \cite{moravec2007compressive} that a sparse 2D image can be uniquely determined (up to global phase) with significantly reduced, undersampled Fourier magnitude measurements. Modified iterative projection algorithms are proposed in \cite{moravec2007compressive,marchesini2008ab,mukherjee2012iterative} for CPR in the noise free case. Convex optimization method is developed in \cite{ohlsson2011compressive} inspired by \cite{candes2011phaselift} with slow computing speed. The state of the art is generalized approximate message passing (GAMP) based algorithm which is introduced in \cite{schniter2012compressive} and shown to have good performance in both recovery accuracy and computing speed.

In this paper, nonconvex $\ell_1$ minimization problems are formulated for noiseless and noisy CPR problems inspired by existing CS results. A desirable property is shown in the widely studied scenario of real-valued, nonnegative image reconstruction that the image of interest is an optimal solution of the formulated $\ell_1$ minimization problem in the noise free case. First-order iterative algorithms are proposed to solve the formulated problems based on the alternating direction method (ADM) \cite{glowinski1980lectures,boyd2011distributed} in convex optimization. Numerical simulations are provided for simulated sparse images and nonnegative image to demonstrate the fast, accurate and robust performance of the proposed approach.

The rest of the paper is organized as follows. Section \ref{sec:theory} presents our problem formulations for the noiseless and noisy CPR problems. Section \ref{sec:algorithm} presents the ADM-based algorithms for the proposed $\ell_1$ minimization problems. Section \ref{sec:simulation} provides the simulation results and Section \ref{sec:conclusion} concludes the paper.

%
%
%
%
%
%

\section{Compressive Phase Retrieval via $\ell_1$ Minimization} \label{sec:theory}
\subsection{Problem Formulations}
We study the recovery of sparse signals from their Fourier magnitude measurements, known as compressive phase retrieval. Inspired by the recent CS technique, we propose to solve the following LASSO-like optimization problem
\equ{\min\lbra{\onen{\m{x}}+\frac{\lambda}{2}\twon{\abs{\cF_{\Omega}\m{x}}-\m{b}}^2}, \label{formu:lasso}}
where $\m{x}$ is the signal of interest, $\cF_{\Omega}$ denotes the discrete Fourier transform (DFT) constrained on the index set $\Omega$, $\m{b}$ denotes the observed (generally noisy) Fourier magnitude measurements on $\Omega$ and $\lambda>0$ is a regularization parameter. The second term in (\ref{formu:lasso}) fits the observed magnitude data while the $\ell_1$ norm is used to promote the signal sparsity. An equivalent formulation of (\ref{formu:lasso}) is
\equ{\min\onen{\m{x}},\st\twon{\abs{\cF_{\Omega}\m{x}}-\m{b}}\leq\epsilon, \label{formu:BPDN}}
where $\epsilon$ controls the fidelity of the reconstruction to the measured data. It is known that (\ref{formu:lasso}) and (\ref{formu:BPDN}) are equivalent with appropriate choices of $\lambda$ and $\epsilon$. As $\lambda\rightarrow+\infty$ and $\epsilon\rightarrow0$, both (\ref{formu:lasso}) and (\ref{formu:BPDN}) reduces to
\equ{\min\onen{\m{x}},\st\abs{\cF_{\Omega}\m{x}}=\m{b} \label{formu:BP}}
in the noise free case.

\subsection{Performance Guarantee}
We note that the $\ell_1$ minimization problem (\ref{formu:BP}) has been studied in \cite{moravec2007compressive}. However, there have been few theoretical results on its performance analysis. The following theorem deals with a practical and widely studied scenario where the signal of interest is nonnegative (nonnegative means real-valued as well in this paper). Although far from satisfactory, this result nevertheless provides some assurance on the reliability of the proposed $\ell_1$ minimization approach.

\begin{thm} If $\m{x}^o\succeq0$ and its Fourier magnitudes $\m{b}=\abs{\cF_{\Omega}\m{x}^o}$ is observed, where $\Omega$ contains the zero frequency, then $\m{x}^o$ is an optimal solution to (\ref{formu:BP}). \label{thm:optimality}
\end{thm}

\emph{Proof:} We need to show that the inequality $\onen{\m{x}}\geq\onen{\m{x}^o}$ holds for any $\m{x}$ satisfying $\abs{\cF_{\Omega}\m{x}}=\abs{\cF_{\Omega}\m{x}^o}$. To do this, consider the DC (zero frequency) term of the frequency data which is proportional to the sum of all elements of the signal vector. So the DC term must be nonnegative since $\m{x}^o\succeq0$. It follows from $\abs{\cF_{\Omega}\m{x}}=\abs{\cF_{\Omega}\m{x}^o}$ that
$\sum_i x_i=\sum_i x^o_i$.
As a result,
\equ{\begin{split}\onen{\m{x}}
&=\sum_i\abs{x_i}\\
&\geq\sum_i x_i=\sum_i x^o_i\\
&=\onen{\m{x}^o},\end{split}}
where the last equality follows from the positivity of $\m{x}^o$.
\BOX


Theorem \ref{thm:optimality} states that a nonnegative signal has the least $\ell_1$ norm among all possible candidates which result in the same Fourier magnitudes. Though our focus is on the recovery of sparse signals, Theorem \ref{thm:optimality} does hold for general nonnegative signals and hence brings new insights to the applicable scope of the $\ell_1$ optimization. It is noted that the proposed $\ell_1$ optimization methods do not result in a unique solution. In fact, they suffer from at least the same ambiguities as existing approaches to conventional phase retrieval. For example, an optimal solution to either one of (\ref{formu:lasso})-(\ref{formu:BP}) after a modification of constant global phase, spatial shift and/or conjugate inversion remains optimal.

\section{ADM for Compressive Phase Retrieval} \label{sec:algorithm}

\subsection{Preliminary: Alternating Direction Method} \label{sec:ADM}
The augmented Lagrangian alternating direction method (ADM) solves the structured optimization problem
\equ{\min_{\m{x},\m{z}}f\sbra{\m{x}}+g\sbra{\m{z}},\st \m{A}\m{x}+\m{B}\m{z}=\m{c}, \label{formu:ADMproblem}}
where $f\sbra{\m{x}}$ and $g\sbra{\m{z}}$ are convex functions of $\m{x}$ and $\m{z}$, respectively. The augmented Lagrangian function of the problem is given by
\equ{\begin{split}
\cL&\sbra{\m{x},\m{z},\m{y}}\\
&=f\sbra{\m{x}}+g\sbra{\m{z}}+\Re\inp{\m{y},\;\m{A}\m{x}+\m{B}\m{z}-\m{c}}\\ &\quad+\frac{\alpha}{2}\twon{\m{A}\m{x}+\m{B}\m{z}-\m{c}}^2\\
&=f\sbra{\m{x}}+g\sbra{\m{z}}+\frac{\alpha}{2}\twon{\m{A}\m{x}+\m{B}\m{z}-\m{c} +\frac{1}{\alpha}\m{y}}^2\\
&\quad-\frac{1}{2\alpha}\twon{\m{y}}^2,\end{split}}
where $\m{y}$ is a Lagrangian multiplier and $\alpha>0$ is a penalty parameter. Starting with $\m{y}^0$ and $\m{z}^0$, the ADM iterates as follows:
\lentwo{\equa{\m{x}^{k+1}
&=& \arg\min_{\m{x}}\cL\sbra{\m{x},\m{z}^{k},\m{y}^k},\label{formu:solvex}\\ \m{z}^{k+1}
&=& \arg\min_{\m{z}}\cL\sbra{\m{x}^{k+1},\m{z},\m{y}^k},\label{formu:solvez}\\ \m{y}^{k+1}
&=& \m{y}^k + \beta\alpha\sbra{\m{A}\m{x}^{k+1}+\m{B}\m{z}^{k+1}-\m{c}}, }
}where $\beta\in\sbra{0,\;\frac{\sqrt{5}+1}{2}}$ guarantees the convergence under some technical assumptions \cite{glowinski1980lectures}. The ADM is very efficient when explicit solutions are available for (\ref{formu:solvex}) and (\ref{formu:solvez}). The ADM has been a popular approach to solutions of large scale problems, since it can typically produce a modestly accurate solution within a few tens of iterations though its convergence to high accuracy may be slow \cite{boyd2011distributed}.

\subsection{ADM for Compressive Phase Retrieval}
Problems (\ref{formu:lasso})-(\ref{formu:BP}) are nonconvex due to the presence of the magnitude operator. In this section we solve (\ref{formu:lasso})-(\ref{formu:BP}) using the ADM. Problem (\ref{formu:lasso}) can be formulated into
\equ{\min\lbra{\onen{\m{x}}+\frac{\lambda}{2}\twon{\abs{\m{z}_{\Omega}}-\m{b}}^2},\st \cF\m{x}-\m{z}=\m{0}.}
According to (\ref{formu:ADMproblem}), we see in this case that $f\sbra{\m{x}}=\onen{\m{x}}$ is convex while $g\sbra{\m{z}}=\frac{\lambda}{2}\twon{\abs{\m{z}_{\Omega}}-\m{b}}^2$ is nonconvex. By recognizing that the DFT is unitary, $\m{x}$ can be explicitly updated using a soft thresholding operator (later given in (\ref{formu:xupdate_lasso})). Let $\m{s}=\cF\m{x}+\frac{1}{\alpha}\m{y}$. Then the augmented Lagrangian involving $\m{z}$ is expressed as
\equ{\begin{split}
&\frac{\lambda}{2}\twon{\abs{\m{z}_{\Omega}}-\m{b}}^2+\frac{\alpha}{2}\twon{\m{z}- \m{s}}^2\\
\geq&\frac{\lambda}{2}\sum_{i\in\Omega}\sbra{\abs{z_i}-b_i}^2 + \frac{\alpha}{2}\sum_{i\in\Omega}\sbra{\abs{z_i}-\abs{s_i}}\\
=&\sum_{i\in\Omega}\frac{\lambda+\alpha}{2}\sbra{\abs{z_i}-\frac{\lambda b_i+\alpha\abs{s_i}}{\lambda+\alpha}}^2 + C, \end{split}}
where the `$=$' in the inequality holds if $\sgn\sbra{z_i}=\sgn\sbra{s_i}$ for $i\in\Omega$ and $z_i=s_i$ for $i\notin\Omega$, and $C$ is a constant independent of $\m{z}$. As a result, the augmented Lagrangian obtaines the minimum at the $\m{z}$ such that $z_i=\frac{\lambda b_i+\alpha\abs{s_i}}{\lambda+\alpha}\sgn\sbra{s_i}$ for $i\in\Omega$ and $z_i=s_i$ for $i\notin\Omega$. So the alternating direction algorithm for (\ref{formu:lasso}) is summarized as follows:
{\lentwo\equa{\m{x}^{k+1}
&=&\bS_{\alpha^{-1}}\sbra{\cF^{-1}\sbra{\m{z}^k-\frac{1}{\alpha}\m{y}^k}},\label{formu:xupdate_lasso}\\ z_i^{k+1}
&=&\left\{\begin{array}{ll}\frac{\lambda b_i+\alpha\abs{s_i^{k+1}}}{\lambda+\alpha}\sgn\sbra{s_i^{k+1}}, & \text{ if } i\in\Omega,\\ s_i^{k+1}, & \text{ otherwise},\end{array}\right.\label{formu:zupdate_lasso}\\ \m{y}^{k+1}
&=&\m{y}^k+\beta\alpha\sbra{\cF\m{x}^{k+1}-\m{z}^{k+1}},\label{formu:yupdate_lasso}}
}where $\m{s}^{k+1}=\cF\m{x}^{k+1}+\frac{1}{\alpha}\m{y}^k$, $\bS_{\alpha}(w)=\sgn\sbra{w}\cdot(\abs{w}-\lambda)^+$ is a soft thresholding operator with $(\cdot)^+=\max(\cdot,0)$.

Next, (\ref{formu:BPDN}) can be formulated as
\equ{\min_{\m{x},\m{z}}\onen{\m{x}}+\bI_{S}\sbra{\m{z}},\st\cF\m{x}-\m{z}=\m{0},}
where $S=\lbra{\m{z}:\twon{\abs{\m{z}_{\Omega}}-\m{b}}\leq\epsilon}$, and $\bI_S$ is an indicator function with $\bI_S\sbra{\m{z}}=0$ if $\m{z}\in S$, and $\bI_S\sbra{\m{z}}=+\infty$ otherwise. As a result, we have $f\sbra{\m{x}}=\onen{\m{x}}$ and $g\sbra{\m{z}}=\bI_{S}\sbra{\m{z}}$ is nonconvex since the set $S$ is nonconvex. Denote $\cP_S$ the projection (in Euclidean norm) onto $S$.
According to the ADM we obtain the algorithm for (\ref{formu:BPDN}) which is the same as that for (\ref{formu:lasso}) except for the update rule of $\m{z}$, $\m{z}^{k+1}=\cP_S\sbra{\m{s}^{k+1}}$. So, provided $\m{b}\succeq\m{0}$ we have
\equ{z^{k+1}_i=\left\{\begin{array}{ll}\sbra{\theta \abs{s^{k+1}_i}+\sbra{1-\theta}b_i}\sgn\sbra{s^{k+1}_i}, & \text{ if } i\in\Omega, \\s^{k+1}_i, & \text{ otherwise},\end{array}\right.\label{formu:zupdate_BPDN}}
where $\theta=\min\sbra{1,\;\frac{\epsilon}{\twon{\abs{\m{s}^{k+1}_{\Omega}}-\m{b}}}}$ and $\m{s}^{k+1}$ is defined as before. A detailed derivation of (\ref{formu:zupdate_BPDN}) is omitted.

As $\lambda\rightarrow+\infty$ and $\epsilon\rightarrow0$, both (\ref{formu:zupdate_lasso}) and (\ref{formu:zupdate_BPDN}) reduce to the following update rule of $\m{z}$ for solving (\ref{formu:BP}):
\equ{z_i^{k+1}=\left\{\begin{array}{ll}b_i\sgn\sbra{s_i^{k+1}}, & \text{ if } i\in\Omega,\\ s_i^{k+1}, & \text{ otherwise},\end{array}\right.\label{formu:zupdate_BP} }
while again the updates for $\m{x}$ and $\m{y}$ are the same as (\ref{formu:xupdate_lasso}) and (\ref{formu:yupdate_lasso}), respectively.

Unfortunately, the proposed ADM algorithms cannot be guaranteed to converge. In fact, the convergence issue remains a difficult problem in the nonconvex phase retrieval problem and a mathematically rigorous analysis has not been found for many existing algorithms for conventional phase retrieval including HIO, HPR and RAAR \cite{luke2005relaxed}. The performance of the proposed algorithms will be numerically studied in Section \ref{sec:simulation}, where we observe that the proposed algorithms can either converge to a good result or not converge at all.

\section{Numerical Simulations} \label{sec:simulation}
\begin{figure}
\centering
  \subfigure[Original image]{
    \label{Fig:satellite}
    \includegraphics[width=1.65in]{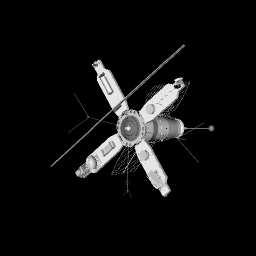}}%
  \subfigure[$M/N=0.2$]{
    \label{Fig:satellite_dot2}
    \includegraphics[width=1.65in]{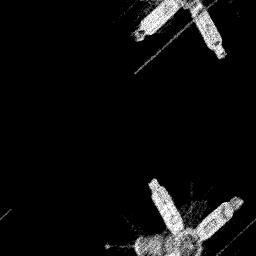}}
  \subfigure[$M/N=0.5$]{
    \label{Fig:satellite_dot5}
    \includegraphics[width=1.65in]{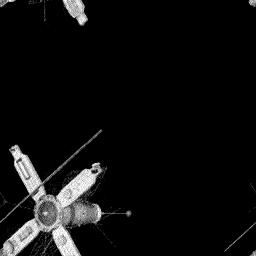}}%
  \subfigure[$M/N=1$]{
    \label{Fig:satellite_1}
    \includegraphics[width=1.65in]{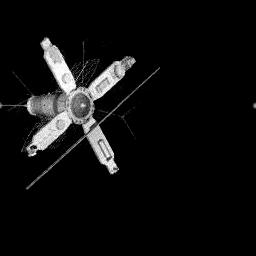}}
\centering
\caption{Results of nonnegative image reconstruction with SNRs of (b) $11.6$, (c) $15.3$ and (d) $18.2$ dB, respectively. The input SNR is $30$ dB.} \label{Fig:result_satellite}
\end{figure}

We provide numerical simulations to demonstrate the performance of the proposed ADM algorithms in this section. 2D Random sparse images and a nonnegative satellite image shown in Fig. \ref{Fig:satellite} are considered. Due to the page limit, a thorough comparison of our method with existing approaches will be reported in a future publication.

\textbf{Algorithm implementation:} The algorithms start with $\m{y}^0=\m{0}$ and a random $\m{z}^0$ such that $\abs{\m{z}_{\Omega}}=\m{b}$ and $z_i=0$ for $i\notin\Omega$. $\alpha$ is set such that a fixed portion, $\rho$, of the elements of $\m{x}^1$ are kept nonzero with $\rho=0.6$. We set $\beta=0.5$ for random sparse images and $\beta=0.8$ for the satellite image. The algorithm is terminated if $\frac{\max\sbra{\alpha\twon{\m{z}^{k}-\m{z}^{k-1}}, \;\cF\m{x}^k-\m{z}^k}}{\twon{\m{b}}}\leq10^{-3}$ or a maximum number of iterations, 500, is reached.

\begin{figure}
\centering
  \includegraphics[width=3.3in]{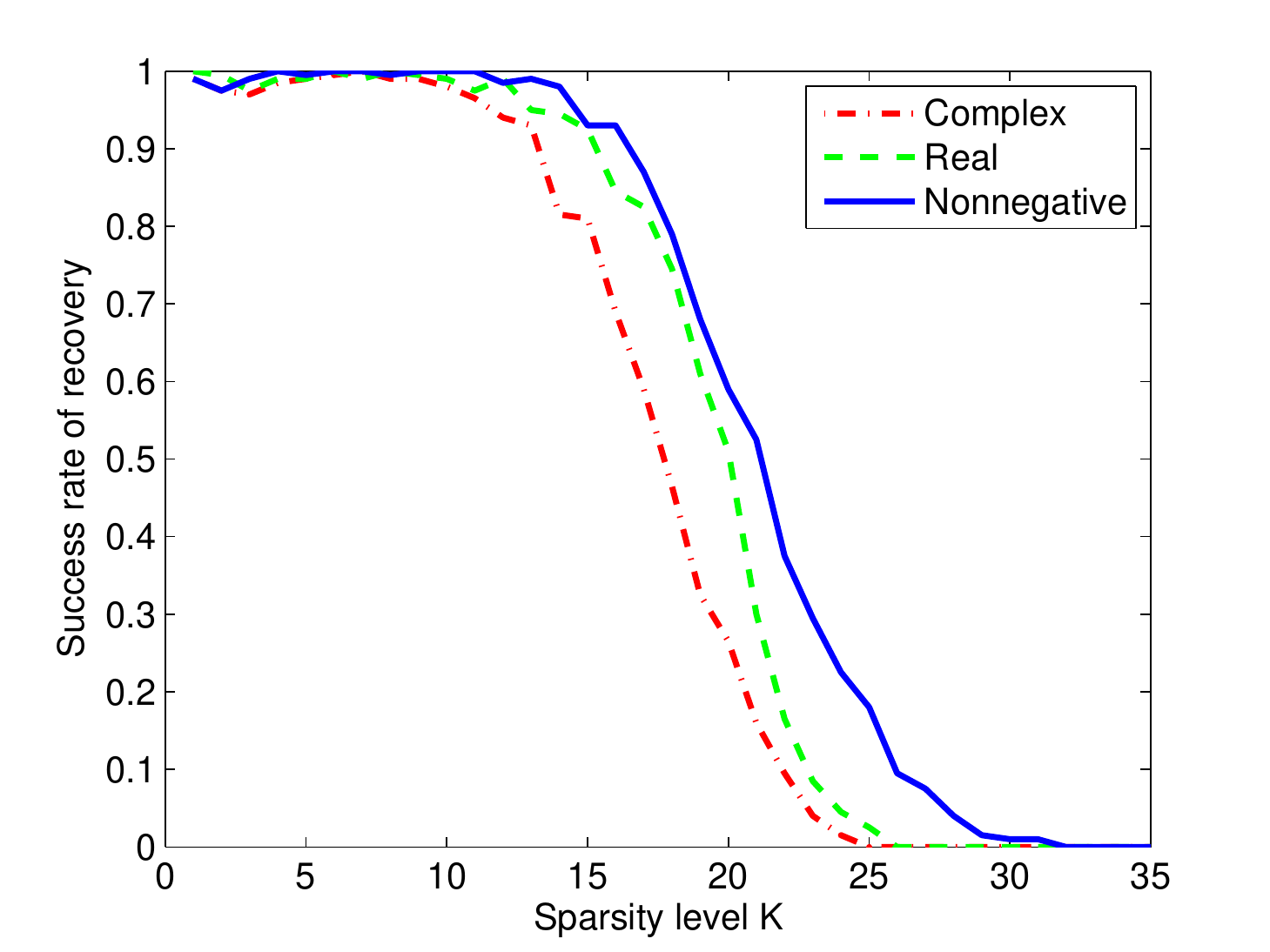}
\centering
\caption{Success rates of recovering complex, real and nonnegative random images with respect to the sparsity level $K$. The number of image pixels and Fourier magnitude measurements are $N=256$ and  $M=128$, respectively.} \label{Fig:successrate_vs_sparsity}
\end{figure}

In the first simulation, we consider random images in the noiseless case and study the success rate of recovery versus the sparsity level $K$. Three types of images are considered including complex, real and nonnegative ones to test whether they have different recovery performance since Theorem \ref{thm:optimality} holds only for nonnegative signals. We consider images of dimension $16\times 16$ (number of image pixels $N=16^2=256$) with image sparsity level $K$ varying from $1$ to $35$ and acquire a number of $M=N/2=128$ random Fourier magnitude measurements. An image is claimed to be correctly recovered if the relative root mean squared error (RMSE) $\frac{\twon{\widehat{\m{x}}-\m{x}^o}}{\twon{\m{x}^o}}\leq10^{-2}$, where $\widehat{\m{x}}$ denotes the recovered image after removing the possible ambiguities of constant global phase, spatial shift and conjugate inversion. For each combination of $K$ and the image type, $200$ trials are repeated. The success rates of recovery are presented in Fig. \ref{Fig:successrate_vs_sparsity}. In the case of a moderate or low sparsity level, the proposed algorithm is observed to converge to the correct solution in most trials starting from a random point. Moreover, it is shown that nonnegative images have the highest success rate of recovery followed by real and then complex ones. The inherent reason will be explored in future studies.

\begin{figure}
\centering
  \includegraphics[width=3.3in]{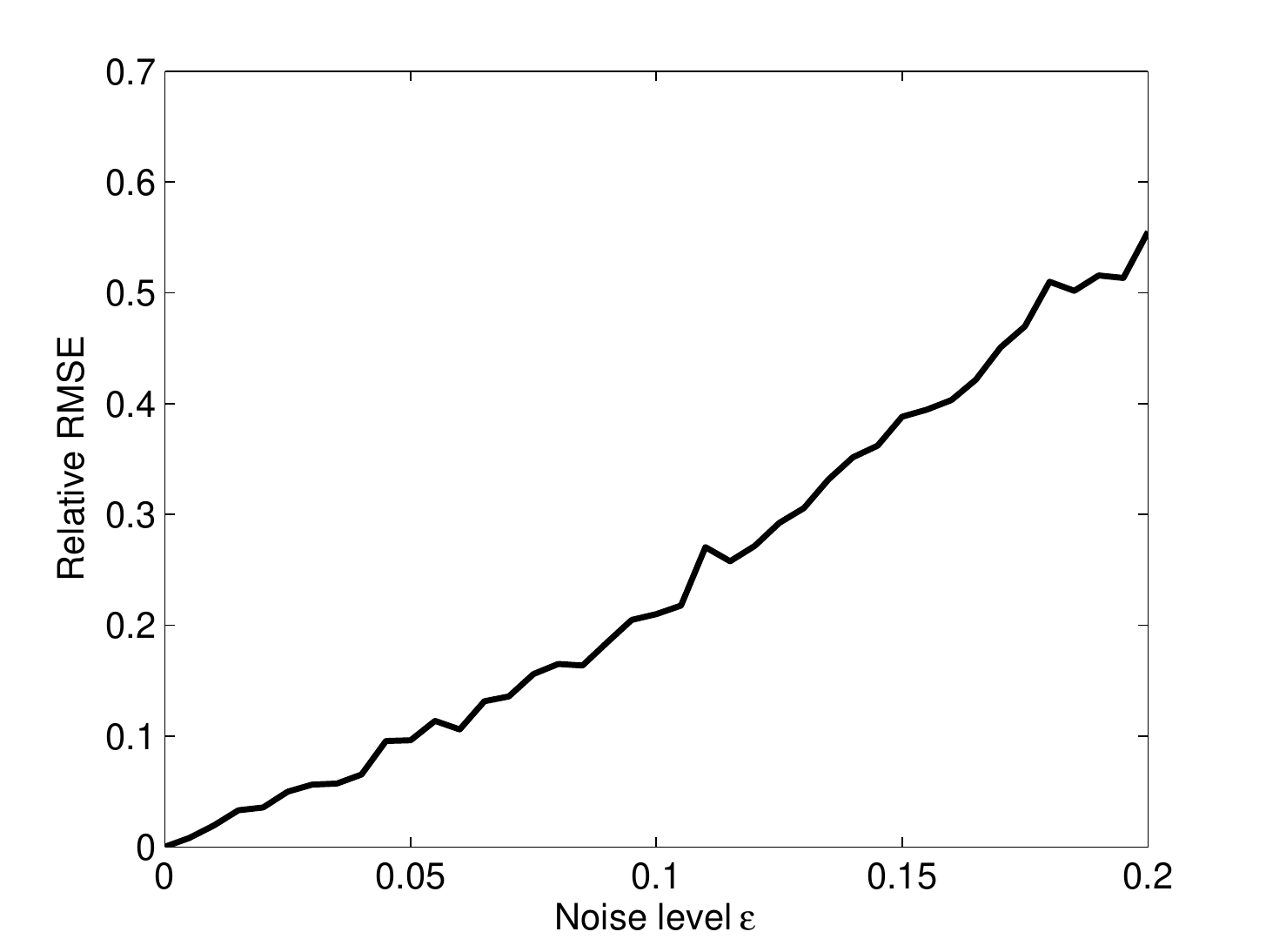}
\centering
\caption{Reconstruction error of random sparse images vs. the noise level $\epsilon$ with $N=256$, $M=128$ and $K=8$.} \label{Fig:RRMSE_vs_noise}
\end{figure}

The second simulation studies the variation of the recovery error with respect to the noise energy. We fix the sparsity level $K=8$ and set $N$ and $M$ as before. Complex random images are scaled to unit Frobenius norm. After obtaining the noiseless Fourier magnitude measurements, a Gaussian random noise is added with the noise energy $\epsilon=\twon{\abs{\cF_{\Omega}\m{x}^o}-\m{b}}$ varying from $0$ to $0.2$ with a step size of $0.005$. The averaged relative RMSE is presented in Fig. \ref{Fig:RRMSE_vs_noise}, where it is shown that the recovery error grows approximately linearly with the noise level $\epsilon$.

The last simulation studies the reconstruction of a nonnegative $256\times 256$ satellite image ($N=256^2=65536$) which is shown in Fig. \ref{Fig:satellite}. The satellite image has a sparsity ratio $K/N\approx0.14$. The number of Fourier magnitude measurements, $M$, is set such that $M/N=0.2,\;0,5$ and $1$, respectively. In each setting, a Gaussian random noise is added to the noiseless magnitude measurements such that the signal to noise ratio (SNR) is $30$ dB. A fixed number of iterations, 200, is used to reconstruct the images. Our simulation results are presented in Fig. \ref{Fig:result_satellite}, where, remarkably, a faithful reconstruction is obtained with only $20\%$ of the Fourier magnitude measurements. Without accounting for the effects of the possible ambiguities mentioned before, the three reconstructed images have SNRs of $11.6$, $15.3$ and $18.2$ dB, respectively. Moreover, the proposed algorithm is very fast and takes $7.8$, $8.8$ and $9.3$ s, respectively, to obtain the reconstructed images using Matlab v7.7.0 on a PC.

\section{Conclusion} \label{sec:conclusion}
The noiseless and noisy compressive phase retrieval problem was studied in this paper while the $\ell_1$ norm is exploited to promote the signal sparsity inspired by compressive sensing. The optimality of the formulated $\ell_1$ minimization problem was proven for nonnegative signals in the sense that the signal of interest is an optimal solution of the $\ell_1$ minimization problem in the noise free case. Efficient alternating direction algorithms were proposed for the problem solving and promising results were presented to demonstrate their performance.

\bibliographystyle{IEEEtran}
\end{document}